\documentclass[shortnote]{jpsj3}
\usepackage{txfonts}
\usepackage{color}
\usepackage{bm}

\title{
Collapsing Behavior of the Ferrimagnetic Ground State of the $S=1/2$ Heisenberg Antiferromagnet on the Lieb Lattice due to Frustration
}

\author{\mbox{Rito Furuchi$^1$,
Hiroki Nakano$^1$,
T$\hat{\rm o}$ru Sakai$^{1,2}$}}
\inst{
$^1$
Graduate School of Science,
University of Hyogo,
Kamigori,
Hyogo 678-1297, Japan \\
$^2$
National Institutes for Quantum and Radiological Science and Technology,
(QST),
SPring-8
Sayo, Hyogo 679-5148, Japan 
} 

\abst{
We study the $S = 1/2$ Heisenberg antiferromagnet on the Lieb lattice accompanied by additional interactions that create frustration. The system exhibits a ferrimagnetic ground state in the absence of frustration. Further, we successfully observe a novel type of collapsing behavior of the ferrimagnetism via numerical diagonalization. The ferrimagnetic state is observed to collapse in a discontinuous manner. Furthermore, different ground states with small spontaneous magnetizations are obtained before the spontaneous magnetization finally disappears.
}

\begin{document}
\maketitle
Ferrimagnetism, which involves both ferromagnetic and antiferromagnetic properties, has been extensively studied since its discovery\cite{ferri}.
The antiferromagnet on the Lieb lattice exhibits ferrimagnetism, which is explained in the Marshall--Lieb--Mattis (MLM) theorem\cite{Marshall,LiebMattis}. If an additional interaction that creates frustration is switched on to the Lieb lattice antiferromagnet, the MLM theorem no longer holds. In previous studies, the collapsing behaviors of the ferrimagnetic ground state of the Lieb lattice antiferromagnet have been examined for various frustrated interactions through numerical diagonalization. These examinations clarified the existence of two types of collapsing behavior. In the first case, the spontaneous magnetization in the ground state continuously decreases and finally disappears \cite{collapsedia,collapsekagome}. Meanwhile, in the second one, the spontaneous magnetization of the ferrimagnetic ground state discontinuously collapses to the ground state without spontaneous magnetization \cite{collapseone}. These behaviors result in the question: Is there another type of collapsing behavior of the ferrimagnetic ground state of the antiferromagnet?

\begin{figure}[htb]
  \begin{center}
    \includegraphics[width=8.0cm]{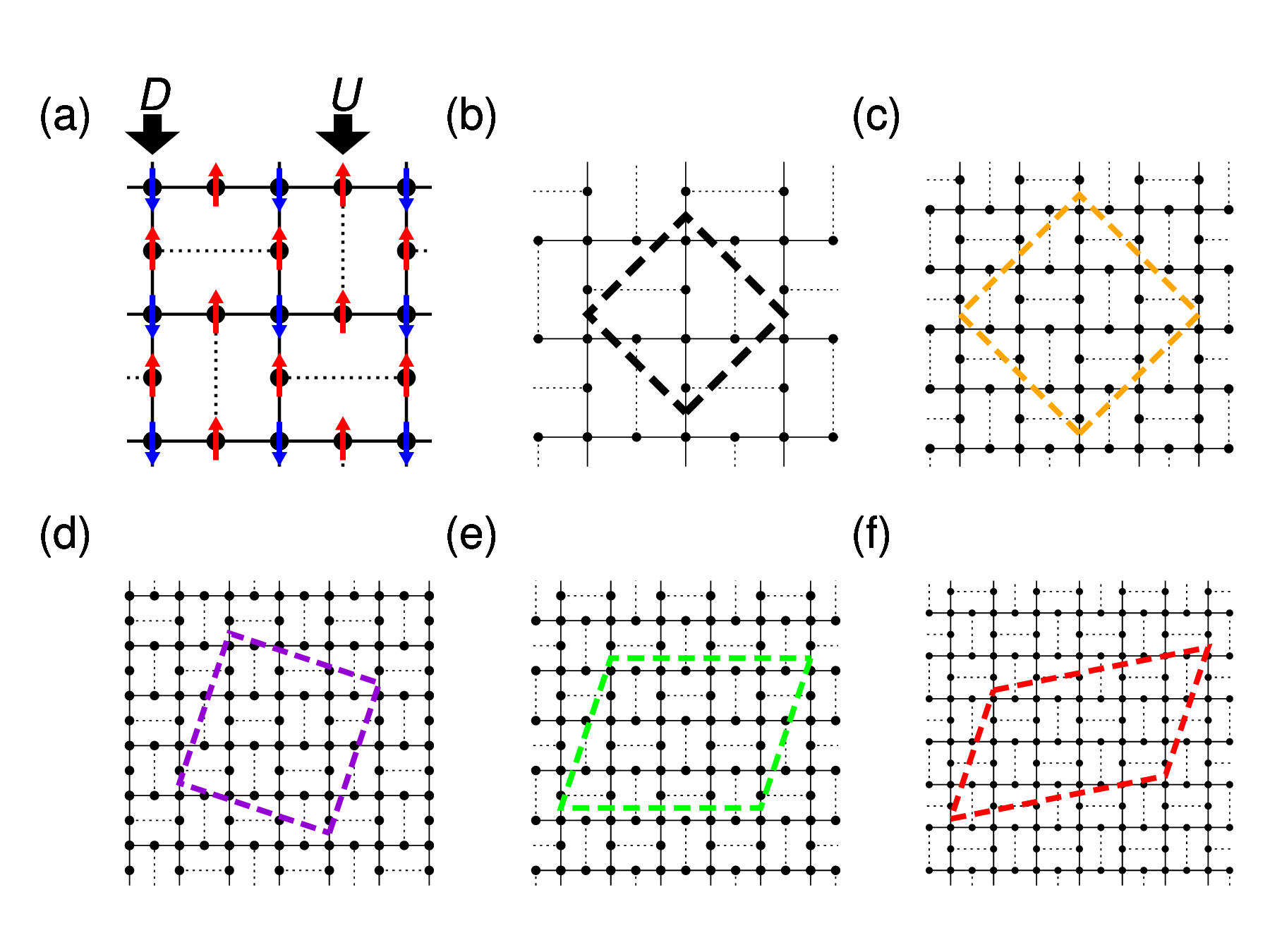}
  \end{center}
  \caption{(Color online)
Lattice investigated in this study. The Lieb lattice is illustrated using closed circles and solid bonds that correspond to spin sites and their networks, respectively. The dotted bonds represent additional interactions that create frustrations.
Panel (a) shows the ferrimagnetic spin state with arrows, 
as well as sublattices $U$ and $D$. 
Panel (b) shows a primitive unit cell via broken thick lines.  
Panels (c), (d), (e), and (f) illustrate finite-size clusters 
via the broken thin lines for $N=$24, 30, 36, and 42, respectively.
}
\label{lat}
\end{figure}

Under these circumstances, the purpose of this study is to clarify a third type of collapsing behavior of the ferrimagnetic ground state of the Lieb lattice antiferromagnet. We investigate the route for changing the case from the Lieb lattice antiferromagnet depicted in Fig.~\ref{lat}. 
We perform numerical diagonalizations for this frustrated case for systems up to a size that has not been treated before. Consequently, our investigation clarifies that the present case exhibits the collapsing behavior of a novel type, namely the third type. 

Let us explain the model Hamiltonian treated in this study.
It is expressed as
\begin{eqnarray}
  \mathcal{H} = \sum_{\langle i \in D ,j \in U\rangle} J\bm{S}_{i} \cdot \bm{S}_{j} 
  + \sum_{\langle i \in U ,j \in U\rangle} J^{\prime}\bm{S}_{i} \cdot \bm{S}_{j}, \label{hamil}
\end{eqnarray}
where $\bm{S}_{i}$ denotes the $S=1/2$ spin operator at site $i$. 
The number of spins is denoted by $N$. 
The interaction bonds of the first term form the Lieb lattice, as illustrated by the solid lines in Fig.~\ref{lat}. 
The spin sites of the Lieb lattice are divided into two sublattices denoted by $U$ and $D$ in Fig.~\ref{lat}~(a). 
The numbers of sites in the primitive unit cell shown in Fig.~\ref{lat}~(b) 
in sublattice $U$ and $D$ are 2 and 1, respectively. 
This difference in the number of sites leads to the occurrence of a ferrimagnetic ground state. The second term of Eq.~(\ref{hamil}) denotes 
additional antiferromagnetic interactions, as illustrated by the dotted lines in Fig.~\ref{lat}. 
This study focuses on the effect of frustration caused by the addition of interactions. We consider the case wherein the amplitudes of the first and second terms, denoted by $J$ and $J^{\prime}$, respectively, are positive. 
Hereinafter, we use the interaction ratio $\eta=J^{\prime}/J$, which 
determines the properties of the target system. 
The ferrimagnetic ground state appears for $\eta=0$, corresponding to the Lieb lattice antiferromagnet. 
For  $\eta=1$, the system corresponds to the Cairo-pentagon lattice antiferromagnet. Cairo-pentagon lattice antiferromagnet has been studied as a frustrated magnet, and experimental and theoretical studies have been conducted \cite{cairo_iron,cairo_qm,cairo_ins,cairo_dim}. The ground state for $\eta=1$ does not reveal spontaneous magnetization; therefore, we focus on the case between $\eta$=0 and 1. 

In this study, we conduct calculations for the above model via numerical diagonalization based on the Lanczos and/or Householder algorithms. Our diagonalization calculations are performed in the subspace characterized by $M$ defined by $\sum_{i}S_{i}^{z}$ to obtain the lowest eigenenergy $E (M)$ for a given $N$. The energies are measured in a unit of $J$.
Note that the saturation magnetization, namely the maximum of $M$, 
is $M_{\rm{sat}}=N/2$. Clusters $N=$24, 30, 36, and 42 are treated as illustrated in Fig.~\ref{lat}~(c), (d), (e), and (f), respectively.
Periodic boundary conditions are imposed on the edges of each cluster. The regular squares are treated for $N=$24 and 30, which capture the two dimensionalities of the system well. However, for clusters of $N=$36 and 42, regular squares cannot be treated; instead, parallelograms are used. Some of Lanczos diagonalizations have been performed using the MPI-parallelized code, which was originally developed for the study of Haldane gaps\cite{haldane}.
The usefulness of our program was confirmed through large-scale parallelized calculations\cite{hnpack}.

\begin{figure}[htb]
  \begin{center}
  \includegraphics[width=8.0cm]{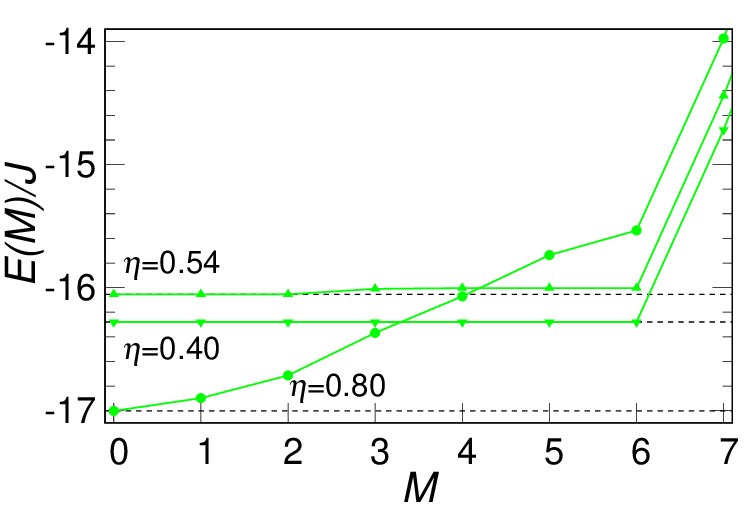}
\caption{(Color online) 
$M$ dependence of the lowest-energy level for various $\eta$.
Results for $N=36$ are presented.
Broken lines indicate the ground-state energy for each case of $\eta$.
}\label{Clu_Com} 
\end{center}
\end{figure}

Now, let us observe the $M$ dependence of the lowest energy $E (M)/J$ in each subspace characterized by $M$. The behavior of the $M$ dependence provides information on the spontaneous magnetization $M_{\rm{spo}}$ for each of $\eta$. The results for the $N=$36 cluster are depicted in Fig~\ref{Clu_Com}.  
For $\eta=0.40$, all of $E (M)$ are numerically identical for the cases from $M=0$ to 6, and $E (M)$ for $M > 6$ increases. This behavior indicates $M_{\rm{spo}}=6$, which corresponds to one-third of $M_{\rm sat}=18$ for $N=36$. This spontaneous magnetization suggests that ferrimagnetism in the Lieb lattice antiferromagnet still survives. For $\eta=0.80$, $E (M=0)$ is lower than $E (M)$ for $M \ge 1$. This behavior suggests that the spontaneous magnetization disappears. Further, for $\eta=0.54$, $E (M)$ from $M=0$ to 2 are numerically identical, and $E (M)$ for $M \ge 3$ becomes larger. This behavior indicates $M_{\rm{spo}}=2$.  This also suggests that the ground state for this $\eta$ is certainly different from the ferrimagnetic ground state of the Lieb lattice antiferromagnet. 

\begin{figure}[htb]
  \begin{center}
  \includegraphics[width=8.0cm]{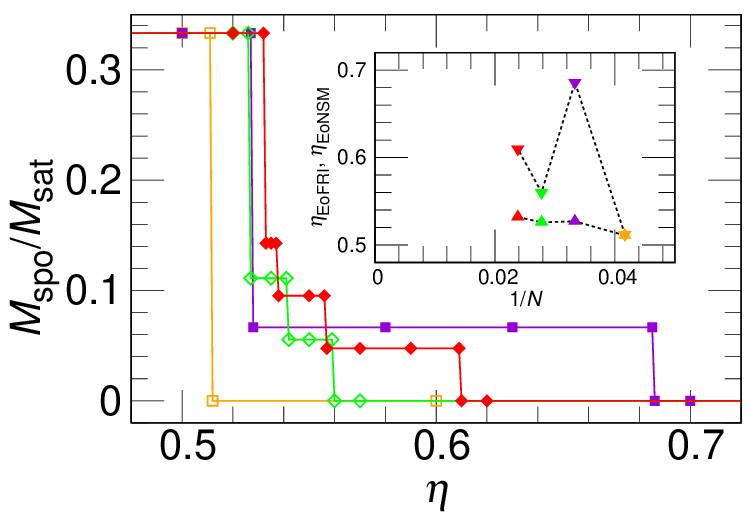}
\end{center}
\caption{(Color online)
Change in the spontaneous magnetization in the ground state of the system. 
Results for $N=42$, 36, 30, and 24 are depicted by red closed diamonds, green open diamonds, violet closed squares, and orange open squares, respectively. 
The inset shows the size dependence of $\eta_{\rm EoFRI}$ and $\eta_{\rm EoNSM}$.
}\label{size_c}
\end{figure}

Next, let us further examine the $\eta$ dependence of spontaneous magnetization for various $N$; results are depicted in the main panel of Fig~\ref{size_c}. Although the cluster of $N=24$ does not reveal the ground state of intermediate $M_{\rm{spo}}$, all the larger clusters clearly exhibit intermediate-$M_{\rm{spo}}$ cases: $M_{\rm{spo}}=1$ for $N=30$, $M_{\rm{spo}}=1$ and 2 for $N=36$, and $M_{\rm{spo}}=1$, 2, and 3 for $N=42$. 
Note that the ferrimagnetic ground states of $N=30$, 36, and 42 correspond to $M_{\rm{spo}}=5$, 6, and 7, respectively. Therefore, the ground states of $M_{\rm{spo}}$ do not appear between $(1/6)M_{\rm{sat}}$ and $(1/3)M_{\rm{sat}}$. This behavior suggests that the ferrimagnetism of the Lieb lattice antiferromagnet discontinuously collapses owing to frustration and that the ground state with almost half of $M_{\rm{spo}}$ of the ferrimagnetism appears. For an even larger $\eta$, $M_{\rm{spo}}$ gradually decreases, and spontaneous magnetization finally disappears. Notably, these behaviors have not been observed in previous studies\cite{collapsedia,collapsekagome,collapseone}.

To capture the behavior of the appearance of the ground state with an intermediate $M_{\rm{spo}}$, we observe the edge of the ferrimagnetic phase denoted by $\eta_{\rm EoFRI}$ and the edge of the phase with no spontaneous magnetization denoted by $\eta_{\rm EoNSM}$. The results of the $1/N$ dependence of $\eta_{\rm EoFRI}$ and $\eta_{\rm EoNSM}$ are depicted in the inset of Fig~\ref{size_c}. As evident, the dependence on the system size is small for $\eta_{\rm EoFRI}$. Our diagonalization data suggest that $\eta_{\rm EoFRI}$ appears to converge at approximately $\eta_{\rm EoFRI}\sim 0.532$ obtained for $N=42$. However, the dependence of $\eta_{\rm EoNSM}$ is significantly larger than that of $\eta_{\rm EoFRI}$. Even for complex dependencies, we do not capture a clear behavior trend indicating that $\eta_{\rm EoNSM}$ approaches $\eta_{\rm EoFRI}$. Thus, the complex dependence of $\eta_{\rm EoNSM}$ may be related to whether the shapes of the treated clusters are regular squares or parallelograms; however, the complexity remains unclear. To clarify the reason and whether the intermediate region between $\eta_{\rm EoFRI}$ and $\eta_{\rm EoNSM}$ survives within the thermodynamic limit, further intensive investigations should be conducted. 

In summary, we have studied the collapse of the ferrimagnetism of the Lieb lattice antiferromagnets owing to the frustration goint to the Cairo-pentagon lattice antiferromagnet using numerical diagonalization. Our results have clarified the appearance of ground states with intermediate spontaneous magnetizations. Because of frustration, the ferrimagnetic state discontinuously collapses at the edge of the ferrimagnetic phase. However, the intermediate states with nonzero but small spontaneous magnetization are stably observed prior to the disappearance of the spontaneous magnetization. This behavior differs from the two collapsing behaviors observed in Refs.~\ref{collapsekagome}, \ref{collapsedia},and \ref{collapseone}.
Further examination will deepen our understanding of the frustration effects.  

\begin{acknowledgment}
This research was partially supported by KAKENHI (grant numbers 20H05274, 20K03866, and 23K11125). Non-hybrid thread-parallel calculations in the numerical diagonalizations were based on TITPACK version 2, coded by Nishimori. We used the computational resources of the supercomputer Fugaku provided by RIKEN through the HPCI System Research projects (Project IDs: hp230114, hp230532, and hp230537). Some of computations were performed using the facilities of the Institute for Solid-State Physics, The University of Tokyo.
\end{acknowledgment}

\end{document}